\documentclass{article}
\usepackage{emulateapj,timesfonts,graphicx}

\makeatletter

\newenvironment{inlinefigure}{%
\def\@captype{figure}%
\noindent\begin{minipage}{0.999\linewidth}\begin{center}}
{\end{center}\end{minipage}\smallskip}
\makeatother

\def\***#1{{\sc #1}}
\def\plan#1{\relax}
\def\Plan#1{\relax}
\def\PLAN#1{\relax}

\renewcommand{\arraystretch}{1.2}

\def\Mx{M_{x}}
\def\Mopt{M_{\rm opt}}

\DeclareSymbolFont{euletters}{U}{eur}{m}{n}
\DeclareMathSymbol{\alpha}{\mathord}{euletters}{'013}
\def\Ka{K{\small$\alpha$}}
\def\Kaone{\Ka$_1$}
\def\Katwo{\Ka$_2$}

\def\cygx1{{Cyg~X-1}}
\def\eV{e\kern-0.08em V}

\begin{document}

\lefthead{VIKHLININ}
\righthead{MASS MEASUREMENT IN BLACK-HOLE X-RAY BINARIES}
\submitted{Submitted to the ApJ Letters, Apr 21, 1999}

\title{A Method of Mass Measurement in Black Hole Binaries \\Using Timing and
  High Resolution X-ray Spectroscopy} 

\author{A. Vikhlinin}
\affil{Harvard-Smithsonian Center for Astrophysics, 60 Garden St.,
  Cambridge, MA 02138;
  avikhlinin@cfa.harvard.edu}

\begin{abstract}
  In X-ray binaries, several percent of the compact object luminosity is
  intercepted by the surface of the normal companion and re-radiated through
  Compton reflection and the K-fluorescence. This reflected emission follows
  the variability of the compact object with a delay approximately equal to
  the orbital radius divided by the speed of light. This provides the
  possibility of measuring the orbital radius and thus substantially
  refining the compact object mass determination compared to using optical
  data alone. We demonstrate that it may be feasible to measure the time
  delay between the direct and reflected emission using cross-correlation of
  the light curves observed near the \Ka\ line and above the K-edge of
  neutral iron.  In the case of \cygx1, the time delay measurement is
  feasible with a $3\times10^5-10^6\,$s observation by a telescope with a
  1000~cm$^2$ effective area near 6.4~k\eV\ and with a $\sim 5\,$\eV\ energy
  resolution.  With longer exposures, it may be possible to obtain mass
  constraints even if an X-ray source in the binary system lacks an optical
  counterpart.
\end{abstract}

\keywords{binaries: spectroscopic---X-rays: general---X-rays: stars}

\section{Introduction}\label{sec:intro}

The existence of stellar-mass black holes in our Galaxy is almost
exclusively established by measuring the mass of the compact object in
several bright X-ray binaries. A compact object more massive than
approximately $3\,M_\odot$ is unstable and should collapse to form a black
hole (Rhoades \& Ruffini 1974, Chitre \& Hartle 1976). Therefore, if the
dynamically measured mass of a luminous X-ray source exceeds $3\,M_\odot$,
it is considered a firm black hole candidate.

In X-ray binaries, the orbital velocity of the compact object cannot be
measured by means of optical spectroscopy. This means that the only directly
measured quantity is the mass function $f(M)$,
\begin{equation}\label{eq:f(M)}
  f(M) = \frac{\Mx^3 \sin^3 i}{(\Mx+\Mopt)^2},
\end{equation}
where $\Mx$ and $\Mopt$ are masses of the compact object and normal
companion, respectively, and $i$ is the inclination angle of the orbit. The
mass function is the lower limit of the compact object mass. In many high
mass X-ray binaries, the mass function is uninterestingly low; for example,
$f(M)=0.25\,M_\odot$ for \cygx1\ (Gies \& Bolton 1986).  A determination of
$\Mx$ is possible if $\Mopt$ and $\sin i$ are independently
constrained. Extensive work has been done to estimate these parameters for
several black hole candidates (Bolton 1972, Liutyj, Sunyaev \& Cherepashchuk
1975, McClintock \& Remillard 1986, Remillard, McClintock \& Bailyn 1992,
Herrero et al.\ 1995, Filippenko, Matheson \& Barth 1995, Bailyn et al.\
1995, Remillard et al.\ 1996).  However, the derived compact object mass can
be rather uncertain due to its strong dependence on $\Mopt$ and especially
$\sin i$.  For \cygx1, mass values in the range 6--16$\,M_\odot$ are found
in the literature (Dolan \& Tapia 1989, Gies \& Bolton 1986).

The mass measurement will be much more accurate if some information about
the orbit of the compact object is available. Unfortunately, very fast
motions of the material in the vicinity of the black hole make it impossible
to measure the orbital velocity from the Doppler shifts of spectral lines.
Instead, we propose to measure the orbital radius using the following
approach. In high mass X-ray binaries, some fraction of the compact object
X-ray luminosity is intercepted by the normal companion surface (the
intercepted fraction in low mass binaries is too low for an application of
our method). A significant fraction of this energy is re-radiated in the
X-ray band either through Compton reflection or line fluorescence.  This
emission reaches the observer with a time delay roughly equal to $a/c$,
where $a$ is the separation between the normal companion and the compact
object.  Since the compact object emission is usually variable, it may be
possible to measure the time delay and thus estimate the orbital radius. We
show below that this measurement can be carried out using temporally
resolved high resolution spectroscopy of the fluorescent \Ka\ line of
neutral iron. Let us demonstrate how this new information refines the
measurement of the compact object mass. Suppose for simplicity that the
orbit is circular and the radius of the normal companion is negligible
compared to the orbital radius.  The time delay at orbital phase $\phi$ is
\begin{equation}\label{eq:delay}
  \Delta t = a/c\left[1-\sin(i\,\cos 2\pi\phi)\right].
\end{equation}
Equations (\ref{eq:f(M)})--(\ref{eq:delay}) together with the Kepler's third
law, 
\begin{equation}\label{eq:kepler-3}
  a^3 = P^2 (\Mx+\Mopt) G/4\pi^2, 
\end{equation}
can be used to express $\Mx$ as a function of just one parameter, $\Mopt$,
$i$, or $a$. In particular, if the time delay measurement is performed at
inferior conjunction ($\phi=0.5$),
\begin{equation}\label{eq:mx-from-delay}
  \Mx=\frac{f^{1/3} \;g^{2/3}}{\sin i\; (1+\sin i)^2} \quad {\rm or} \quad
  \Mx=\frac{(\Mx+\Mopt) f^{1/3}}{g^{1/3}-(\Mx+\Mopt)^{1/3}},
\end{equation}
where $g=4\pi^2 (c\Delta t)^3/P^2G$. If the time delay is measured at
several orbital phases, all system parameters can be determined
independently. Moreover, useful mass constraints are possible in this case
even if no optical data are available (\S~\ref{sec:discussion}).

\section{Measurement strategy}\label{sec:strategy}

The X-ray emission reflected from the surface of the normal companion
consists of the Compton-reflected continuum and the fluorescent emission.
While the reflected continuum is weak compared to the direct emission from
the black hole, the fluorescent line emission can be relatively strong. The
strongest fluorescent line is the iron \Ka\ doublet (Basko, Sunyaev \&
Titarchuk 1974). Basko (1978) estimates the equivalent width of $\approx
7\,$\eV\ for this line in \cygx1. With $\approx 5\,$\eV\ spectral resolution
of X-ray calorimeters, it should be possible to measure the source light
curve in a narrow spectral interval around \Kaone\ and \Katwo, where the
fluorescent line contributes $\approx 50\%$ of the total flux.  The time
delay can be measured by cross-correlating the light curve in the \Ka\ 
region with the continuum light curve which is dominated by the direct black
hole emission.

\Ka\ lines are primarily excited through photoelectric absorption in the
K-edge ($E_K=7.1$~k\eV\ for neutral iron). Therefore, the energy band just
above 7.1~k\eV\ is best for the continuum light curve measurement.  The
exact choice of the energy boundaries is unimportant as long as the source
variability patterns do not change with energy, as is the case in \cygx1\ 
(Nowak et al.\ 1999). Below, we assume that the continuum is measured in the
7.1--9.1~k\eV\ energy band.

The most challenging part of the time delay measurement is determining the
light curve in the \Ka\ line region. This line consists of two components,
\Kaone\ and \Katwo, with energies $E_{\alpha_1}=6.404$ and
$E_{\alpha_2}=6.391$~k\eV\ (Bambynek et al.\ 1972). Both components have
approximately equal natural widths of 3.5~\eV\ (FWHM). The Doppler
broadening of the lines due to thermal motions of the emitting atoms $\Delta
E_{\rm th} = 0.1\,(T/10^5\,{\rm K})^{1/2}$~\eV\ is much smaller than the
natural width of the lines.  Therefore, more than 90\% of the total \Ka\ 
line flux is within a narrow, 20~\eV, energy interval around $E=6.398$~k\eV.
Selecting photons in this energy interval is easy with the 5~\eV\ (FWHM)
energy resolution of future X-ray calorimeters. An approximately 20~\eV\ 
energy interval seems optimal for the time delay measurement. A wider band
would include a larger contribution of the direct continuum without
increasing the line flux significantly.  Narrower intervals would miss flux
in the line wings.

The cross-correlation function of the continuum and line light curves should
contain two prominent peaks. A strong narrow peak around the delay $\tau=0$
corresponds to the auto-correlation of the emission coming directly from the
black hole.  Reflection from the normal companion results in a secondary
broad peak in the correlation function. This peak should have a centroid at
$\tau\approx -a/c$ and width $\Delta\tau\approx R/c$, where $a$ and $R$ are
the radii of the orbit and normal companion, respectively.  The amplitude of
the secondary peak should be proportional to the equivalent width of the
reflected \Ka\ line and the ${\it rms}^2$ of the source flux variability.
Therefore, the time delay measurement is best performed when the reflected
line is strong (i.e.\ near inferior conjunction, Basko 1978), and when the
source is highly variable.

In the next section, we simulate an observation of the time delay in \cygx1\ 
which could be performed by a telescope with an effective area of
1000~cm$^2$ in the 6--9~k\eV\ energy range. Such an instrument approximately
corresponds to one spacecraft of the proposed \emph{Constellation-X}
mission.\footnote{See http://constellation.gsfc.nasa.gov/ for information}

\section{Simulated Observation}

We adapt the binary system parameters from Herrero et al.\ (1995) --- a
radius of the normal companion $R=17\,R_\odot$, a separation between the
components $a=40\,R_\odot$, and an inclination angle $i=35^\circ$. For
simplicity, we assume that the orbit is circular and the normal companion is
spherical. For a similar geometry, Basko (1978) predicted the equivalent
width of the reflected iron line at inferior conjunction
$W_{\phi=0.5}=7$~\eV. Basko assumed Solar chemical composition of the normal
companion. However, some X-ray observations indicate that the iron abundance
may be twice Solar (Kitamoto et al.\ 1984).  Such an overabundance of iron
would increase the fluorescent line flux to $W\approx 14$~\eV. A narrow iron
\Ka\ line of similar amplitude was indeed observed by the \emph{ASCA} SIS
(Ebisawa et al.\ 1996).  We adopt $W_{\phi=0.5}=10\,$\eV, which agrees with
both theoretical predictions and current observations.

As was discussed in \S~\ref{sec:strategy}, we assume that the \Ka\ light
curve is measured in the 20~\eV\ energy interval centered at $6.398$~k\eV,
and the continuum light curve is observed in the 7.1--9.1~k\eV\ band. The
continuum spectrum is assumed to be a power law with a photon index
$\Gamma=1.65$ (Liang \& Nolan 1984). We also allow for the possible presence
of a broad iron line originating in the vicinity of the black hole, for
example, in the outer part of the accretion disk. We assume that the
equivalent width of the broad line is 0.1~k\eV\ (see, e.g., Dove et al.\ 
1998) and that the line width is 0.08~k\eV\ (FWHM) which corresponds to the
velocity dispersion of 2000~km~s$^{-1}$.

For the source variability, we adopt the shot noise model of Lochner et al.\ 
(1991). This model describes the variability of \cygx1\ as a constant
component plus a sequence of exponential flares with durations distributed
in the 0.01--3.22~s interval. The variability level during the 1997
\emph{HEAO~1}~A-2 observation modeled by Lochner et al.\ was quite typical
of the source (35\% \emph{rms}). The shot noise model is adequate for
simulation of the direct emission from the black hole. The light curve of
the reflected radiation is the convolution of the direct emission light
curve with the reflection response to the instantaneous flare. This response
function is calculated below.

\begin{inlinefigure}
  \medskip
  \centerline{\includegraphics{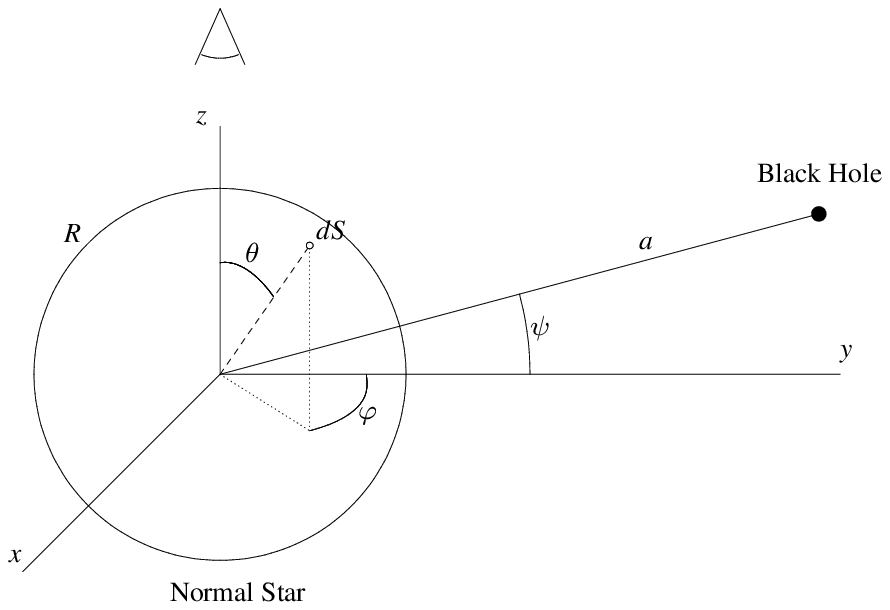}}
  \caption{Geometry of the reflection problem. The line of sight is along the
    $z$-axis. Black hole is in the $y-z$ plane; $\psi=-i\cos 2\pi\phi$,
    where $\phi$ is the orbital phase.}\label{fig:geometry}
\end{inlinefigure}

\begin{figure*}
  \newlength{\figwidth}
  \setlength{\figwidth}{\textwidth}
  \addtolength{\figwidth}{-\columnsep}
  \setlength{\figwidth}{0.5\figwidth}
  
  \begin{minipage}[t]{\figwidth}
    \mbox{}
    \vskip -18pt
    \centerline{\includegraphics[width=\linewidth]{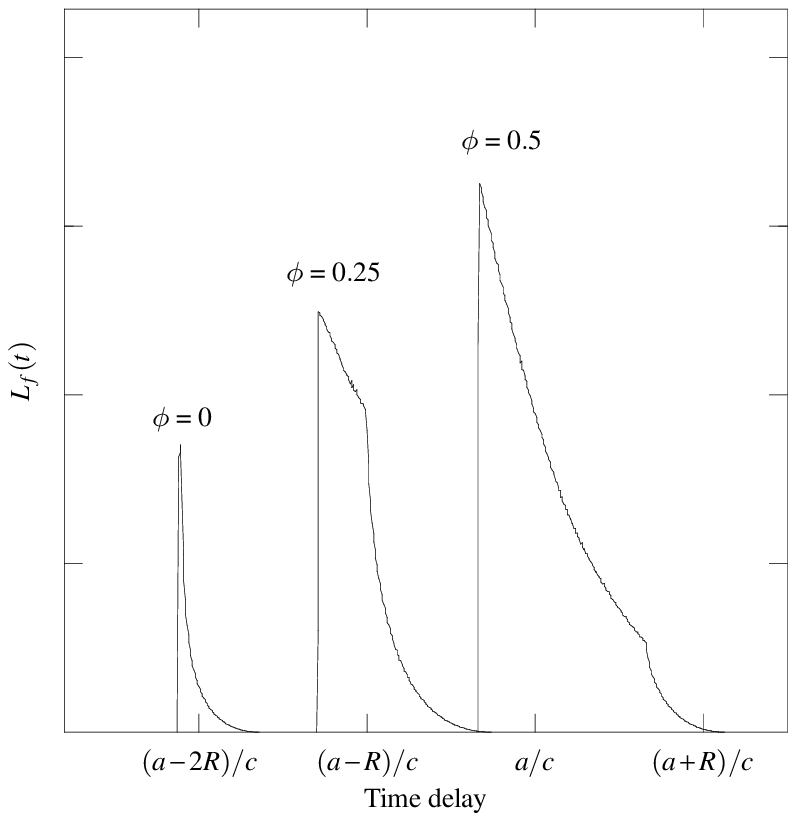}}
    \vskip -7pt
    \caption{Time response of the fluorescent emission in the \cygx1\
      system ($R/c=40\,$s and $a/c=93\,$s in this
      case).}\label{fig:response}
  \end{minipage}
  \hfill
  \begin{minipage}[t]{\figwidth}
    \mbox{}
    \vskip -18pt
    \centerline{\includegraphics[width=\linewidth]{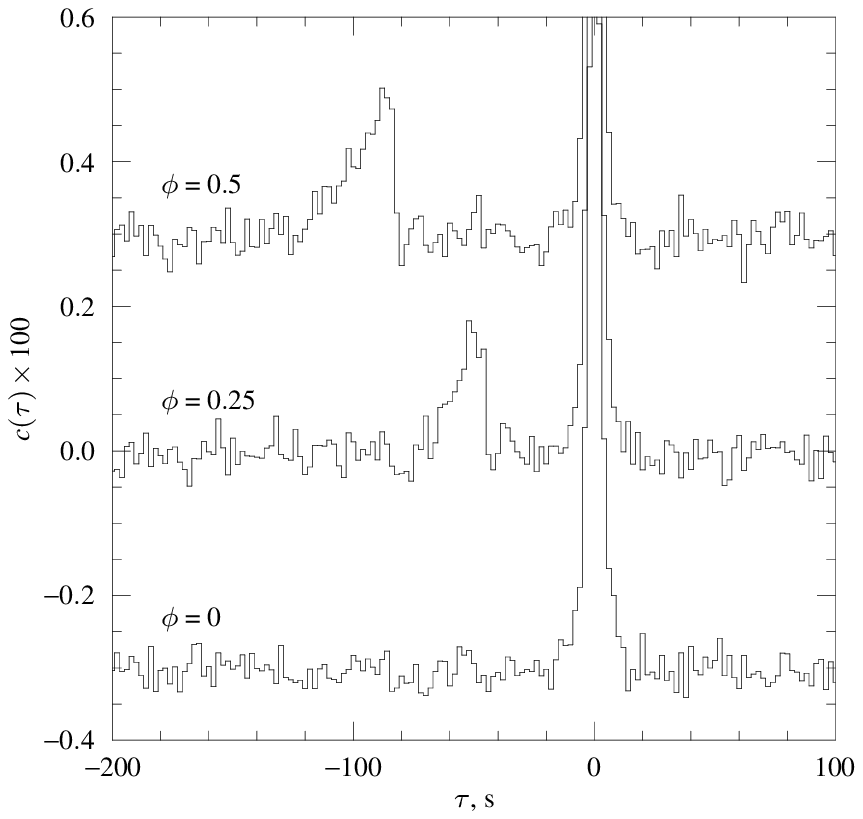}}
    \vskip -7pt
    \caption{The cross-correlation analysis of simulated 1000~ks observations of
      \cygx1\ at three orbital phases. Correlation functions for $\phi=0.5$
      and $\phi=0$ are offset for clarity by $0.003$ and $-0.003$,
      respectively.}\label{fig:cf}
  \end{minipage}
\end{figure*}

Consider the fluorescent emission coming to the observer from the normal
companion surface element $dS$ (Fig.~\ref{fig:geometry}). The distance
between the surface element and the black hole is
\begin{equation}\label{eq:dist}
  l = \left[R^2+a^2-2Ra\,(\cos\psi\sin\theta\cos\varphi+\sin\psi\cos\theta)
  \right]^{1/2}. 
\end{equation}
The reflected emission reaches the observer with a delay $\tau$
\begin{equation}
  c\tau = l +a\sin\psi-R\cos\theta
\end{equation}
relative to the direct emission from the black hole.  The fraction of the
X-ray luminosity intercepted by the surface element is
\begin{equation}
  dL = L \;dS\,\cos\Theta_i /4\pi l^2,
\end{equation}
where $\Theta_i$ is the incident angle of the X-rays,
\begin{equation}\label{eq:thetai}
  \cos\Theta_i = (\cos\psi\sin\theta\cos\varphi+\sin\psi\cos\theta)\,a/l-R/d.
\end{equation}
The ratio of intensities of the observed \Ka\ line and the incident
continuum above the K-edge is a strong function of $\Theta_i$ and the
fluorescence angle, $\theta$. This ratio, $Y(\Theta_i, \theta)$, can be
calculated analytically following the work of Basko. It was also derived by
George \& Fabian (1991) using Monte-Carlo simulations.  George \& Fabian
present $Y(\Theta_i, \theta)$ for an incident power-law spectrum with
$\Gamma=1.7$ on a grid of $\Theta_i$ and $\theta$ values. We use numerical
interpolation of the George \& Fabian results in our calculations below.

The light curve of the fluorescent emission from the surface element  is
\begin{equation}\label{eq:resp:lc}
  f(t) = Y(\Theta_i, \theta)\;  \delta(t-\tau) \;dL
\end{equation}
Integration of equation (\ref{eq:resp:lc}) over the normal companion's
surface provides the time response of the reflected emission.
Figure~\ref{fig:response} shows the results for the \cygx1\ system at three
orbital phases. A different set of responses is expected for a non-spherical
normal companion, for example, the one filling its Roche lobe.  However, the
calculations in this case should be analogous to the procedure outlined
above.

Our simulations proceed as follows. The continuum light curve in the
7.1--9.1~k\eV\ band is simulated using the shot noise model of Lochner et
al. The reflected \Ka\ line light curve is the convolution of the
7.1--9.1~k\eV\ light curve with the time response from
Fig~\ref{fig:response}.  The \Ka\ line flux is rescaled so that the line
equivalent width (relative to the $\Gamma=1.65$ power law continuum) is
10~\eV\ at $\phi=0.5$. The 6.388--6.408~k\eV\ flux as a sum of the \Ka\ line
flux and the contribution of the direct emission calculated as the
7.1--9.1~k\eV\ light curve scaled according to the spectral model which
consists of the power law and a broad 6.4~k\eV\ line (see above).  Finally,
we scale the simulated light curves to reproduce the observed \cygx1\ flux
at 7~k\eV\ (0.05~ph~s$^{-1}$~cm$^{-2}$~k\eV$^{-1}$, Ebisawa et al.\ 1996),
multiply them by the effective area of 1000~cm$^2$, and add Poisson noise.

The cross-correlation functions derived from the simulated $1000\,$ksec
observations are presented in Fig~\ref{fig:cf}. The secondary peak due to
reflected emission is clearly visible at orbital phases $\phi=0.5$ and 0.25.
For $\phi=0$, the secondary cross-correlation peak is weak and not separated
from the autocorrelation of the direct emission (the average delay is only
14~s for the adopted system parameters).

\section{Discussion}\label{sec:discussion}

We have shown that the time delay between the direct and reflected emission
can be measured in a \cygx1\ like system.  Arguably, this measurement
requires long observations. For a convincing detection, a 250~ksec
observation on a telescope with 1000~cm$^2$ effective area around 6~k\eV\ is
required. A longer, 1000~ksec, exposure is needed for a quality measurement
of the cross-correlation function (Fig~\ref{fig:cf}).  These exposures are
longer than the 5.6 day orbital period of \cygx1, and therefore the
observation needs to be split into several intervals, each performed at the
same orbital phase. The required exposure is shorter for larger area
telescopes. Unfortunately, the gain is smaller than a factor of $({\rm
  area})^{1/2}$ because the internal source variability becomes the dominant
source of noise. To achieve a detection significance similar to that
in Fig.~\ref{fig:cf}, $\approx 300$~ksec exposure is a minimum even for very
large area telescopes.

In addition to the fluorescence from the surface of the normal companion,
iron \Ka\ lines can form in the accretion disk or stellar wind. These
emissions are contaminants for the time delay measurement. Fortunately, it
may be possible to separate them in the energy or time domains from the
normal component fluorescence. First, lines originating in the accretion
disk or stellar wind are likely to be significantly broadened because the
reflecting material moves at high speed (a possible detection of the
broadened iron line in Cyg X-1 is reported by Done \& Zycki 1999). Thus, we
can exclude a large fraction of the disk and possibly wind fluorescent
emission by selecting a narrow range around the \Ka\ line rest frame energy.
Second, iron in both wind and disk is likely to be partially ionized because
of the low matter density and high ionizing flux from the black hole. \Ka\ 
lines for ionization states higher than ${\rm Fe}\,{\rm XV}$ are separated
from the ${\rm Fe}\,{\rm I}$ lines by more than 20~\eV\ (House 1969). Third,
the accretion disk or stellar wind fluorescence should produce
cross-correlation functions very different from those in Fig.~\ref{fig:cf}.
Since the outer disk radius is only a small fraction of the black hole Roche
lobe size (Shapiro \& Lightman 1976), the time delay of the fluorescent disk
emission is much shorter than that in Fig.~\ref{fig:cf}. In the case of
stellar wind fluorescence, time delays should be distributed in the broad
range $\Delta\tau\approx r/c$ around the average value
$\overline{\tau}=r/c$, where $r$ is the characteristic radius of the
fluorescence region.  Therefore, a very broad feature in the
cross-correlation function is expected due to stellar wind fluorescence.
Such a feature should be easily distinguishable from the relatively narrow
peak arising from the normal companion reflection. To conclude, the only
consequence of the presence of the iron K-fluorescence from the stellar wind
or accretion disk is to add noise to the cross-correlation function in the
time delay range of interest.

The quality of the data in our simulated observation is sufficient to
determine both the average delay and \emph{shape} of the cross-correlation
peak. The average delay provides the orbital separation between the
components which can be used to refine the black hole mass estimates
(\S~\ref{sec:intro}). The width of the correlation peak is proportional to
the radius of the normal companion. Its shape depends on the shape of the
companion surface and the inclination angle (e.g., the $\phi=0.25$ curve in
Fig.~\ref{fig:response} corresponds to all orbital phases in an $i=0$
system). The shift of the average delay as a function of the orbital phase
also depends on the inclination angle and the companion radius
(eq.~\ref{eq:delay}, Fig.~\ref{fig:response}).  Therefore, a much more
detailed modeling than simple considerations outlined in
eqs.~(\ref{eq:delay})--(\ref{eq:mx-from-delay}) might be possible with
high-quality time delay data. This should result in the compact object mass
determination without the $\sin i$ or $\Mopt$ degeneracies of
eq~(\ref{eq:mx-from-delay}).

Basko (1978) pointed out that the orbital modulation of the reflected \Ka\ 
line flux can be used to estimate the orbital parameters, in particular, the
inclination angle. Our method is independent of Basko's, because we do not
use the line flux. The interpretation of the orbital modulation of the \Ka\ 
line flux can be uncertain. For example, the fluorescence in the stellar
wind reduces the orbital modulation of the line flux; a similar behavior is
expected for low inclination angles (see Basko's Fig.~4). On the contrary,
our method is not directly affected by either stellar wind or accretion disk 
fluorescence.

The time delay technique opens the possibility of mass determination in
X-ray systems without known optical counterparts, such as bright black hole
candidates in the Galactic center region or LMC. In these systems, the
binary period and the epoch of zero orbital phase can be determined by
periodic variations of either the iron line intensity (Basko 1978) or
location of the cross-correlation peak. Near the orbital phase $\phi=0.25$,
the cross-correlation function is independent of the inclination angle and
can be used to measure the orbital radius, $a$, and the radius of the normal
companion, $R$, (Fig.~\ref{fig:response}). The total system mass $\Mopt+\Mx$
is then determined from the Kepler's third law. Under the assumption that
the normal companion almost fills its Roche lobe, which should be the case
in X-ray luminous systems, the ratio $R/a$ defines $\Mx/\Mopt$ and thus
provides an estimate of $\Mx$.

\acknowledgements
This work was supported by the Harvard-Smithsonian Center for Astrophysics
postdoctoral fellowship.

\end{document}